\documentclass[prl,showpacs,twocolumn,preprintnumbers,
amsmath,groupedaddress,amssymb,superscriptaddress,floatfix]{revtex4}
\usepackage[pdftex]{graphicx}
\usepackage[latin1]{inputenc}
\usepackage{xcolor,bm}
\usepackage{nicefrac}

\begin{document}
\draft

\hyphenation{a-long}

\title{Charge doping versus disorder in CeFeAsO: do the in- and
out-of-plane dilutions play the same role?}


\author{G.~Prando}\email[E-mail: ]{g.prando@ifw-dresden.de}
\affiliation{Leibniz-Institut f\"ur Festk\"orper- und
Werkstoffforschung (IFW) Dresden, D-01171 Dresden, Germany}
\author{O.~Vakaliuk}
\affiliation{Leibniz-Institut f\"ur Festk\"orper- und
Werkstoffforschung (IFW) Dresden, D-01171 Dresden, Germany}
\author{S.~Sanna}
\affiliation{Dipartimento di Fisica and Unit\`a CNISM di Pavia,
Universit\`a di Pavia, I-27100 Pavia, Italy}
\author{G.~Lamura}
\affiliation{CNR-SPIN and Universit\`a di Genova, I-16146 Genova,
Italy}
\author{T.~Shiroka}
\affiliation{Laboratorium f\"ur Festk\"orperphysik,
ETH-H\"onggerberg, CH-8093 Z\"urich, Switzerland}\affiliation{Paul
Scherrer Institut, CH-5232 Villigen PSI, Switzerland}
\author{P.~Bonf\`a}
\affiliation{Dipartimento di Fisica and Unit\`a CNISM di Parma,
Universit\`a di Parma, I-43124 Parma, Italy}
\author{P.~Carretta}
\affiliation{Dipartimento di Fisica and Unit\`a CNISM di Pavia,
Universit\`a di Pavia, I-27100 Pavia, Italy}
\author{R.~De Renzi}
\affiliation{Dipartimento di Fisica and Unit\`a CNISM di Parma,
Universit\`a di Parma, I-43124 Parma, Italy}
\author{H.-H.~Klauss}
\affiliation{Institut f\"ur Festk\"orperphysik, Technische
Universit\"at Dresden, D-01062 Dresden, Germany}
\author{S.~Wurmehl}
\affiliation{Leibniz-Institut f\"ur Festk\"orper- und
Werkstoffforschung (IFW) Dresden, D-01171 Dresden, Germany}
\author{C.~Hess}
\affiliation{Leibniz-Institut f\"ur Festk\"orper- und
Werkstoffforschung (IFW) Dresden, D-01171 Dresden, Germany}
\author{B.~B\"uchner}
\affiliation{Leibniz-Institut f\"ur Festk\"orper- und
Werkstoffforschung (IFW) Dresden, D-01171 Dresden,
Germany}\affiliation{Institut f\"ur Festk\"orperphysik, Technische
Universit\"at Dresden, D-01062 Dresden, Germany}

\widetext

\begin{abstract}
We provide direct experimental evidence for the identical effect of
the in-plane Fe$_{1-x}$Co$_{x}$ and of the out-of-plane
O$_{1-x}$F$_{x}$ chemical dilutions on the itinerant
spin-density-wave (SDW) magnetic phase in CeFeAsO. Remarkably, the
suppression of SDW is not sensitive at all to the different kinds of
disorder introduced in the two cases. Still, it is clearly shown
that the sizeable in-plane disorder induced by the
Fe$_{1-x}$Co$_{x}$ substitution is highly effective in suppressing
$T_{\textrm{c}}$. Differently from what is observed in
CeFeAsO$_{1-x}$F$_{x}$, the ordered magnetic phase of the Ce
sublattice is preserved throughout the whole phase diagram in
CeFe$_{1-x}$Co$_{x}$AsO ($x \leq 0.2$). An intriguing effect is
encountered, whereby the magnetic coupling among Ce$^{3+}$ ions is
enhanced by the superconducting phase.
\end{abstract}


\date{\today}

\maketitle

\narrowtext

The substitution of $\sim 10$\% of O$^{2-}$ ions by F$^{-}$ leads to
the emergence of high-$T_{\textrm{c}}$ superconductivity (SC) in
LaFeAsO \cite{Kam08,Lue09} and in other oxy-pnictides
\cite{Zha08,Dre09,San09,Shi11}. Although such chemical dilution is
realized out of the FeAs layers, it is known to induce a charge
doping in the Fe bands. This strategy allows one to reach
$T_{\textrm{c}} \simeq 55$ K in SmFeAsO$_{1-x}$F$_{x}$, namely one
of the highest values among all the Fe-based SC
\cite{Ren08,Joh10,Pra11,Mae12}. In spite of the lower
$T_{\textrm{c}}$ values, also other kinds of chemical substitutions
which still give rise to SC in Fe-based materials are widely
studied. These involve the dilution of transition metal (TM)
elements on the FeAs layers \cite{Ni10,Tan11}, the most common one
being probably cobalt \cite{Sef08a,Pra09,Zha10,Sha13,Sef08b,Yam09}.
Whether the chemical substitution of Fe by other TM elements
effectively induces a charge doping similarly to the case of
O$_{1-x}$F$_{x}$ has been subject of recent intense efforts on both
the computational and the experimental sides, particularly in the
case of Fe$_{1-x}$Co$_{x}$. Contrasting results have been reported
from density-functional theory (DFT) calculations
\cite{Wad10,Ber12a} as well as from measurements via x-rays
absorption \cite{Bit11,Mer12} and photoemission \cite{Lev12}
spectroscopies. Moreover, the perturbation induced by the
Fe$_{1-x}$TM$_{x}$ substitution is expected to introduce a high
degree of in-plane disorder, in turn destabilizing the magnetic
itinerant spin-density-wave (SDW) phase \cite{Wad10,Bon12}. This is
consistent with data for different TM elements in $122$, where a
common scaling for the suppression of SDW is realized by considering
the amount of dilution rather than the effective charge doping
\cite{Ni10}. Such picture clearly goes far beyond the scenario of
the weakening of the Fermi-surface nesting induced by charge doping
\cite{Liu08}. The interplay of the effects of such disorder on the
itinerant magnetic phase and SC is even reported to lead to a
counterintuitive enhancement of $T_{\textrm{c}}$ with increasing the
degree of disorder in the underdoped region of the phase diagram
\cite{Fer12}.

In this Letter we provide direct evidence for the equivalence of the
in-plane Fe$_{1-x}$Co$_{x}$ and out-of-plane O$_{1-x}$F$_{x}$
chemical dilutions in CeFeAsO as far as the magnetic properties are
concerned. Measurements of muon-spin spectroscopy ($\mu^{+}$SR) show
how both these substitutions lead to a \emph{quantitatively
identical} suppression of the SDW phase as a function of $x$. Two
different regimes are detected and a short-range magnetic phase is
found to nanoscopically coexist with SC within a narrow range of $x$
values. Remarkably, the SDW transition temperature $T_{\textrm{N}}$
is not affected at all by the degree of the in-plane disorder.
However, the in-plane disorder itself strongly affects the
properties of SC, $T_{\textrm{c}}$ being sizeably suppressed by the
Fe$_{1-x}$Co$_{x}$ dilution with respect to what is realized via the
O$_{1-x}$F$_{x}$ substitution. The magnetic ordering temperature
$T_{\textrm{N}}^{\textrm{Ce}}$ for the Ce sublattice is preserved
throughout the whole phase diagram in CeFe$_{1-x}$Co$_{x}$AsO
independently from the SDW phase, at variance with what is reported
for CeFeAsO$_{1-x}$F$_{x}$ \cite{Shi11,Mae12}. Finally, an
intriguing correlation among the dome-like features of the phase
diagram for both $T_{\textrm{c}}$ and $T_{\textrm{N}}^{\textrm{Ce}}$
is described suggesting an enhancement of the magnetic coupling
among Ce$^{3+}$ ions mediated by the SC phase.

Powder samples of CeFe$_{1-x}$Co$_{x}$AsO ($0 \leq x \leq 0.2$) were
synthesized as described in detail in the supplemental material
(SM). The characterization of the samples was performed using
electrical transport \cite{Hes09} and dc magnetometry, while a
detailed investigation of $T_{\textrm{N}}$ was performed by means of
$\mu^{+}$SR \cite{Yao11}. The results of the characterizations as
well as the precise information about the definitions of the
critical temperatures $T_{\textrm{N}}$, $T_{\textrm{c}}$ and
$T_{\textrm{N}}^{\textrm{Ce}}$ are reported in the SM.

\begin{figure}[t!]
\begin{center}
\includegraphics[scale=0.2]{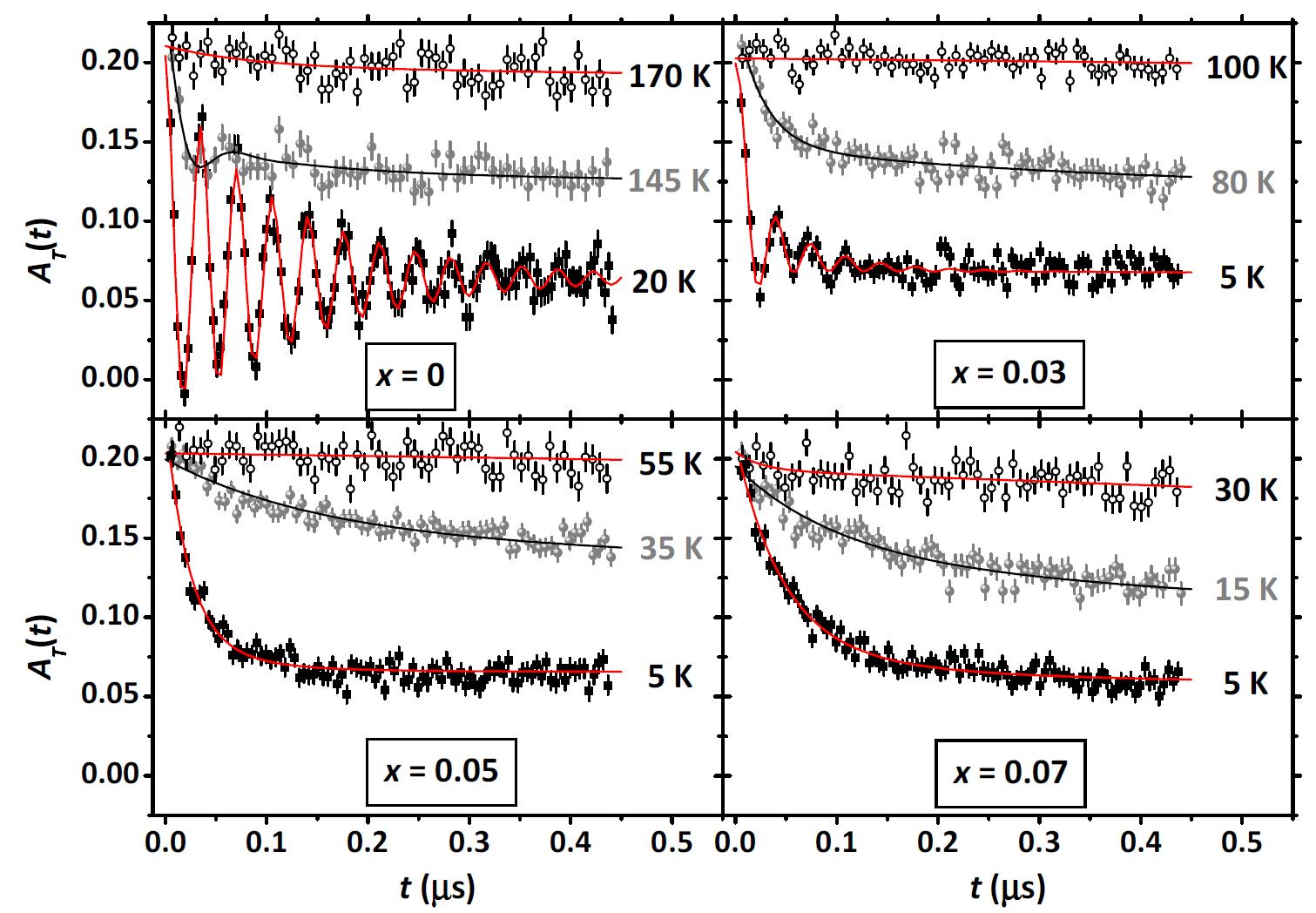}%
\caption{\label{GraDepolarizationAll}(Color online) ZF
$\mu^{+}$-depolarization curves for four CeFe$_{1-x}$Co$_{x}$AsO
samples ($0 \leq x \leq 0.07$) at representative $T$ values.
Continuous lines are best fits to experimental data according to the
analysis described in the SM. Data for $x = 0$ are taken from Ref.
\cite{Shi11}.}
\end{center}
\end{figure}
Sets of experimental $\mu^{+}$ spin-depolarization curves $A_{T}(t)$
in zero-magnetic field (ZF) at different temperatures ($T$) are
shown in Fig.~\ref{GraDepolarizationAll} for several
CeFe$_{1-x}$Co$_{x}$AsO samples ($0 \leq x \leq 0.07$). Data are
analyzed as described in the SM. Two inequivalent crystallographic
sites for $\mu^{+}$ are observed in agreement with previous results
on oxy-pnictides \cite{Mae09,DeR12,Pra13a}. Because of the dipolar
interaction between $\mu^{+}$ and the antiferromagnetic FeAs layers,
$\mu^{+}$ are very sensitive to magnetic ordering even when the
magnetic coherence length $\xi_{\textrm{m}}$ is of the order of few
nanometers. Typically, the precessions of $\mu^{+}$ around the local
magnetic fields are rapidly overdamped when $\xi_{\textrm{m}}
\lesssim 10$ lattice constants \cite{Yao11,San10a}. This is exactly
the case of CeFe$_{1-x}$Co$_{x}$AsO for $x \gtrsim 0.035$, as shown
in Fig.~\ref{GraDepolarizationAll}, a result indicating that the
long-range magnetic order (LRO) realized at $x = 0$, where
$\xi_{\textrm{m}} \gg 1$ nm, is gradually driven to a short-range
magnetic order (SRO), where $\xi_{\textrm{m}} \sim 1$ nm for higher
$x$ concentrations.

\begin{figure}[t!]
\begin{center}
\includegraphics[scale=0.2]{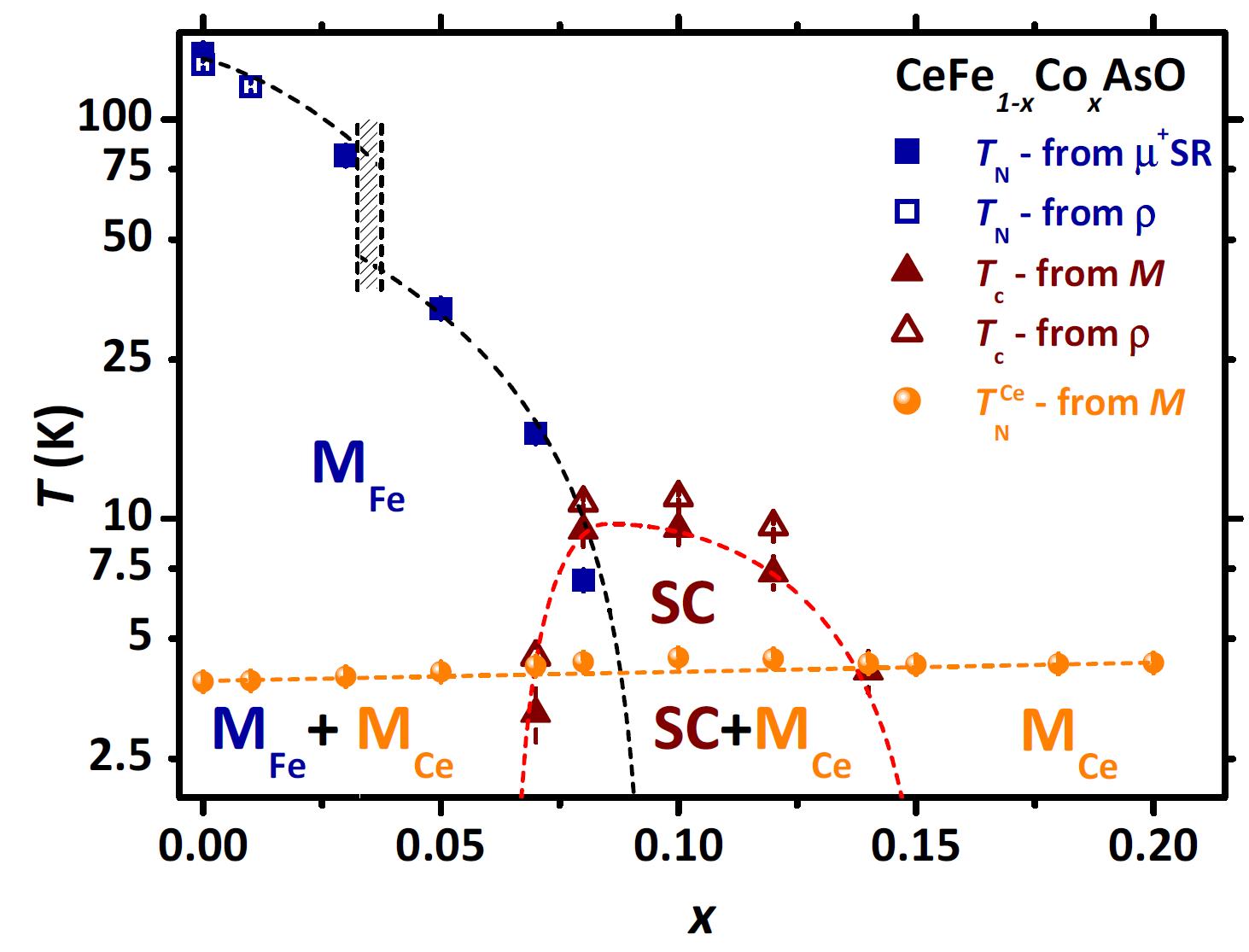}%
\caption{\label{GraPDmagn}(Color online) Electronic phase diagram
for CeFe$_{1-x}$Co$_{x}$AsO. M$_{\textrm{Fe}}$ and M$_{\textrm{Ce}}$
refer to magnetic phases of Fe and Ce, respectively. The dashed
lines are empirical guides for the eyes. Dashed lines on the
magnetic side of the phase diagram are discussed in more detail in
Fig.~\ref{GraPDmagnOnly}.}
\end{center}
\end{figure}
The full electronic phase diagram of CeFe$_{1-x}$Co$_{x}$AsO is
presented in Fig.~\ref{GraPDmagn}. By comparing these results with
what is reported for CeFeAsO$_{1-x}$F$_{x}$ \cite{Shi11}, it is
immediately clear that the SC properties are affected differently by
the two chemical dilutions. In particular, the typical values
$T_{\textrm{c}} \simeq 10$ K for CeFe$_{1-x}$Co$_{x}$AsO at optimal
doping are much lower than what is observed for
CeFeAsO$_{1-x}$F$_{x}$, namely $T_{\textrm{c}} \simeq 35$ K
\cite{Shi11,Mae12,Sha13}. This suggests that the Fe$_{1-x}$Co$_{x}$
dilution is generally detrimental to SC since it introduces a
sizeable degree of disorder directly on the active FeAs layers.
However, the region of coexistence between SDW and SC phases is
similar to what was reported for SmFeAsO$_{1-x}$F$_{x}$ and
CeFeAsO$_{1-x}$F$_{x}$ \cite{San09,Dre09,Shi11}. In particular, an
extremely narrow coexistence region for the two electronic ground
states is observed for both the in-plane and out-of-plane
substitutions in the $1111$ family. Under these conditions, $100$\%
of the sample volume is magnetic while, at the same time, a sizeable
bulk region displays SC (see SM). From these arguments, a
segregation of the two phases at the nanoscopic level was deduced in
SmFeAsO$_{1-x}$F$_{x}$ and CeFeAsO$_{1-x}$F$_{x}$ and the same
scenario holds also in the case of Fe$_{1-x}$Co$_{x}$ dilution. This
scenario is completely different from what is detected in La-based
$1111$ materials \cite{Lue09,Kha11,Pra13b}. It should be stressed
that such nanoscopic coexistence is detected in
CeFeAsO$_{1-x}$F$_{x}$ and CeFe$_{1-x}$Co$_{x}$AsO only when the
magnetic phase is short-ranged. Remarkably, this is qualitatively
different from what is obtained in the electron-doped $122$
materials like Ba(Fe$_{1-x}$Co$_{x}$)$_{2}$As$_{2}$, where the
$x$-range of coexistence is typically much wider \cite{Ni10,Fer10},
superconductivity being reported to emerge still in the presence of
clear coherent oscillations for the SDW phase \cite{Ber12b}. The
appearance of superconductivity even in the presence of a LRO
magnetic phase is possibly a fingerprint of the $122$ family, at
variance with what is generally observed in $1111$ materials. This
can be observed by comparing the results presented in Ref.
\cite{Ber12b} with what is reported in the case of hole-doping
realized by means of the Ba$_{1-x}$K$_{x}$ substitution
\cite{Wie11}.

\begin{figure}[t!]
\begin{center}
\includegraphics[scale=0.2]{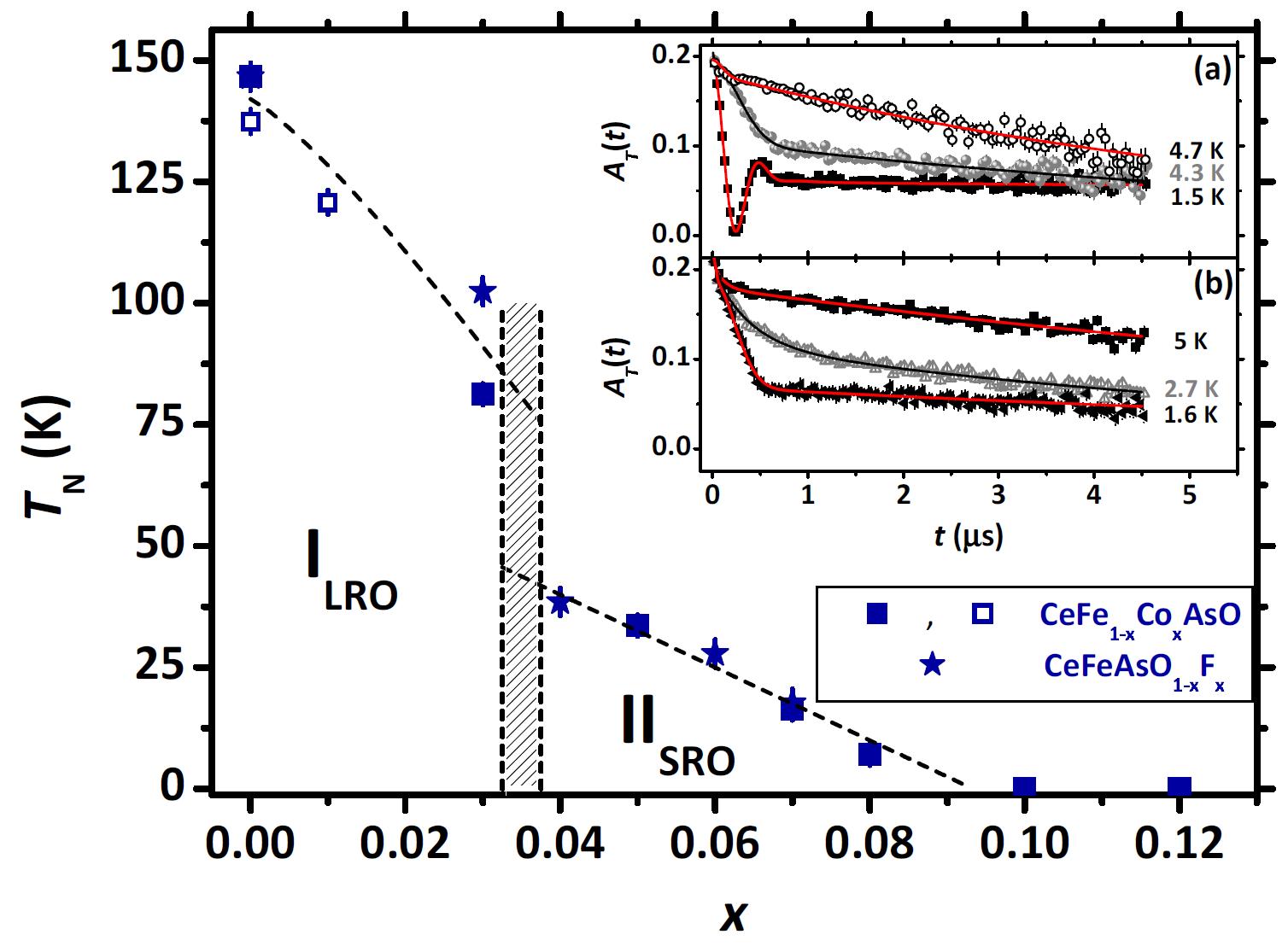}%
\caption{\label{GraPDmagnOnly}(Color online) Main panel:
$T_{\textrm{N}}$ versus $x$ for Co- and F-doped CeFeAsO (the latter
data are taken from Ref. \cite{Shi11}). Close and open symbols refer
to estimates from $\mu^{+}$SR and resistivity, respectively. The
shaded region indicates the crossover between the LRO and SRO
magnetic regimes. The dashed lines are empirical guides for the
eyes. Inset: comparison between ZF $\mu^{+}$-depolarization curves
at representative $T$ for samples in the SC region of the phase
diagram for both CeFe$_{1-x}$Co$_{x}$AsO ($x = 0.12$), panel (a),
and CeFeAsO$_{1-x}$F$_{x}$ ($x = 0.08$), panel (b). Continuous lines
are best fits to experimental data according to the analysis
described in SM.}
\end{center}
\end{figure}
With the aim to precisely compare the effects of the in-plane and
out-of-plane dilutions in CeFeAsO, the values of $T_{\textrm{N}}$
versus $x$ in CeFe$_{1-x}$Co$_{x}$AsO are reported in the main panel
of Fig.~\ref{GraPDmagnOnly}, together with data for
CeFeAsO$_{1-x}$F$_{x}$, the latter being taken from Ref.
\cite{Shi11}. It should be stressed that $x$ concentrations for the
latter materials should be considered as real after an \emph{a
posteriori} determination by means of nuclear magnetic resonance
\cite{Shi11}. Remarkably, results for both the families of compounds
display a common behavior. Magnetism is indeed driven only by the
amount of dilution regardless of the possible different
modifications both of the details of the band structure or of the
introduced degree of in-plane disorder. Two qualitatively different
trends for the suppression of $T_{\textrm{N}}$ with increasing $x$
can be detected, which are associated with LRO and SRO (denoted as
I$_{\textrm{LRO}}$ and II$_{\textrm{SRO}}$ in
Fig.~\ref{GraPDmagnOnly}, respectively). In particular, a linear
dependence of $T_{\textrm{N}}$ on $x$ common to both F- and
Co-diluted compounds is encountered in the intermediate-doping
region close to the appearance of SC. The results are similar to
what is realized in Fe$_{1+y}$Se$_{x}$Te$_{1-x}$, where the region
II is associated with the emergence of a glassy-like magnetism
\cite{Kat10,Lam13}. Remarkably, the phase diagram is also highly
reminiscent of hole-doped cuprates, such as
La$_{2-x}$Sr$_{x}$CuO$_{4}$ and
Y$_{1-x}$Ca$_{x}$Ba$_{2}$Cu$_{3}$O$_{6+y}$, where the glassy
magnetism is also generally reported to be fully suppressed for $x
\sim 0.1$ \cite{Jul03,San10b}. However, the coherent oscillations of
the $\mu^{+}$SR signal are clearly detected also well-inside region
II both for Fe$_{1+y}$Se$_{x}$Te$_{1-x}$ and for the cuprates due to
clusters whose magnetic moment freezes at low $T$
\cite{San10b,Lam13}. This inhomogeneous magnetic order is often
referred to as cluster spin glass.

In spite of the above similarities, clear qualitative differences
among the effects of in-plane and out-of-plane disorders are
enlightened concerning the features of the magnetic phase of the Ce
sublattice. As reported in Refs. \cite{Shi11} and \cite{Mae12},
$T_{\textrm{N}}^{\textrm{Ce}}$ is strongly suppressed in
CeFeAsO$_{1-x}$F$_{x}$ as $x$ increases and eventually vanishes in
correspondence with the full suppression of the SDW phase. A similar
phenomenology was recently reported in CeFe$_{1-x}$Ru$_{x}$AsO even
though a much weaker dependence on $x$ was detected in this case
\cite{Wan12}. This would suggest that the two magnetic phases are
intimately connected and that the ordering of the Ce sublattice is
induced by the molecular field generated by the SDW \cite{Mae12}.
The current results clarify that this is not really the case, since
data for CeFe$_{1-x}$Co$_{x}$AsO unambiguously show that the
magnetic phase of the Ce sublattice is still present at $x = 0.2$,
although the SDW is fully suppressed already at $x = 0.1$ (see
Figs.~\ref{GraPDmagn} and \ref{GraPDmagnOnly}). Similar results were
recently reported also in the case of isovalent dilution in
CeFeAs$_{1-x}$P$_{x}$O \cite{Jes12}. This observation for samples
well inside the SC phase is in agreement with what is reported in
Ref. \cite{Sha13} and it is here reinforced by $\mu^{+}$SR data (see
Fig.~\ref{GraPDmagnOnly}, inset). One clearly detects coherent
oscillations in the case of CeFe$_{0.88}$Co$_{0.12}$AsO indicative
of a long-range ordered phase, even though the transverse damping is
quite severe ($\lambda_{1}^{\textrm{Tr}} \simeq 4$ $\mu$s), see
Fig.~\ref{GraPDmagnOnly}(a). On the other hand, in
CeFeAsO$_{0.92}$F$_{0.08}$ (nominal composition) the ordering of
Ce$^{3+}$ ions can be detected from a qualitative change in the
damping of the transverse relaxation, which from Lorentzian turns
into a markedly Gaussian shape at the lowest $T$ value, see
Fig.~\ref{GraPDmagnOnly}(b). No clear signs of coherent oscillations
can be distinguished in this case, again confirming that the
O$_{1-x}$F$_{x}$ dilution strongly affects the magnetic phase of the
Ce$^{3+}$ ions.

\begin{figure}[t!]
\begin{center}
\includegraphics[scale=0.22]{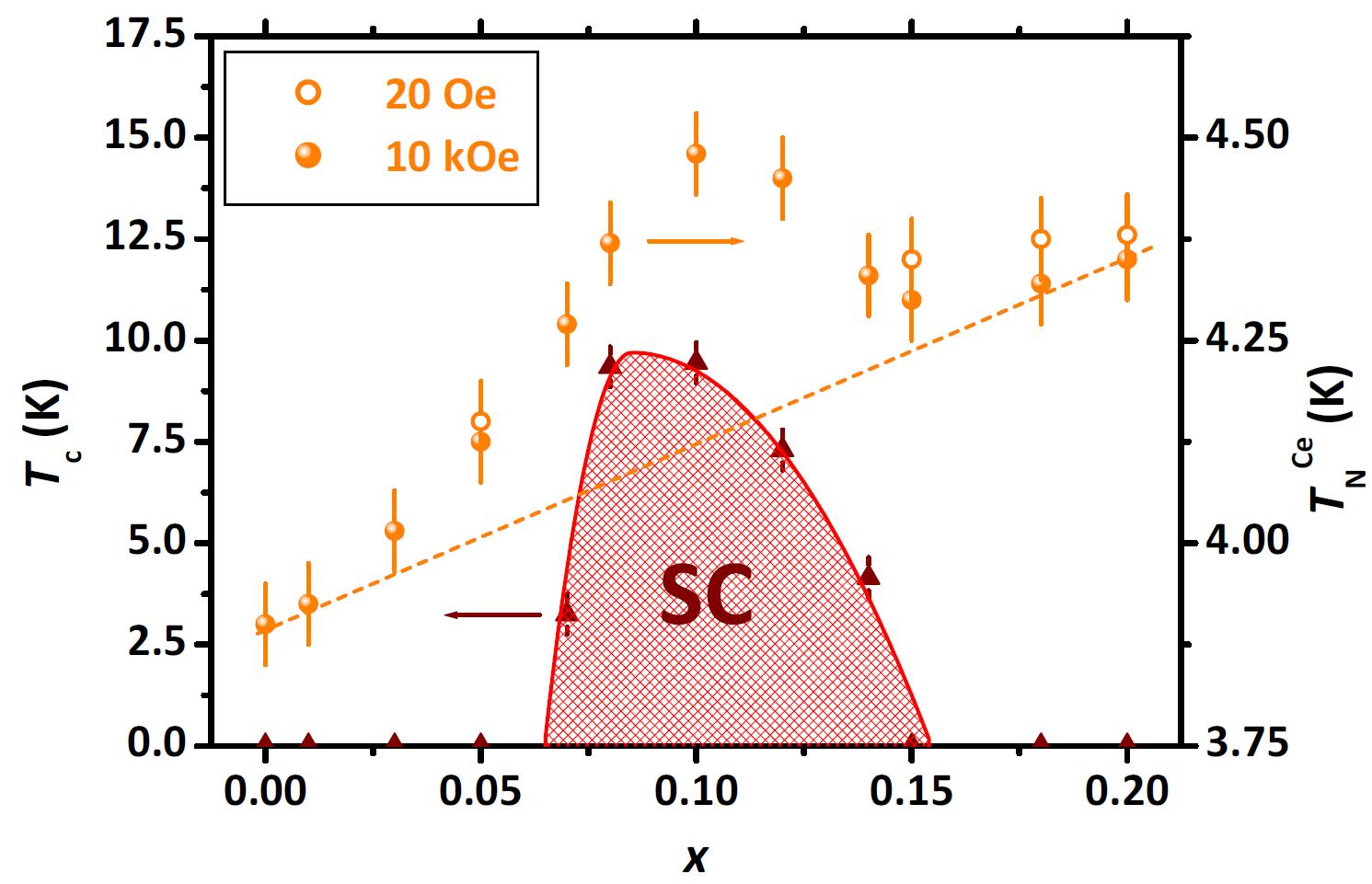}%
\caption{\label{GraPDscANDce}(Color online) Main panel:
$T_{\textrm{c}}$ values as reported after measurements of dc
magnetometry. $T_{\textrm{N}}^{\textrm{Ce}}$ is also plotted (on a
different scale) for two different values of magnetic field. The
continuous and dashed lines are empirical guides for the eyes.}
\end{center}
\end{figure}

These results can be interpreted in the light of two different
mechanisms contributing to the magnetic coupling among Ce$^{+}$
ions, one of direct nature (e.g., superexchange via O$^{2-}$) and
the other of indirect RKKY-like nature, via carriers belonging to
the FeAs layers \cite{Pra10,Akb13}. The strong sensitivity of
$T_{\textrm{N}}^{\textrm{Ce}}$ to the local perturbation realized by
the O$_{1-x}$F$_{x}$ dilution shows that the former mechanism has
the strongest impact to the overall magnetic coupling. Naively, this
is favored by the spatial extension of the Ce$^{3+}$ electronic
orbitals, typically larger than those of other rare-earth ions. The
sizeable sensitivity of the direct term to the increasing F content
completely hinders the modification of the RKKY interaction realized
by the variation of the charge density on the FeAs bands and of the
relative band structure. On the other hand, the in-plane
Fe$_{1-x}$Co$_{x}$ dilution does not affect the CeO layers locally
and the main modification is expected to involve the indirect term.
Moreover, the change of the extent of the $f-d$ hybridization among
Ce$^{3+}$ ions and FeAs bands, triggered by the reduction of the
volume of the crystallographic cell with increasing $x$
\cite{Sha13}, should also be considered. In this respect, it should
be noticed that the gradual increase of
$T_{\textrm{N}}^{\textrm{Ce}}$ upon increasing $x$ in
CeFe$_{1-x}$Co$_{x}$AsO (see the dashed straight line in
Fig.~\ref{GraPDscANDce}) is in qualitative agreement with the
similar trend observed after $\mu^{+}$SR measurements in undoped
CeFeAsO under applied hydrostatic pressure \cite{DeR12}. These
results deserve a more detailed computational investigation in order
to precisely unveil the details of such complex interplay of subtle
many-body effects.

Quite surprisingly, the magnetic coupling among Ce$^{3+}$ ions is
anomalously enhanced in the presence of SC. This is clearly
evidenced in Fig.~\ref{GraPDscANDce}, where a dome-like behaviour
for $T_{\textrm{N}}^{\textrm{Ce}}$ versus $x$ in correspondence with
the emergence of the SC dome is superimposed to the overall linear
increase described above. $T_{\textrm{N}}^{\textrm{Ce}}$ deduced by
means of dc magnetometry is known to be strongly suppressed as the
value of the measuring field increases for typical values $H \gtrsim
10$ kOe (see SM). At first sight, the enhancement of
$T_{\textrm{N}}^{\textrm{Ce}}$ could be associated with the
shielding of the external field by the SC-induced supercurrents.
This effect would apparently increase the critical temperature
itself. However, this scenario is ruled out by the tiny discrepancy
observed among measurements at $H = 20$ Oe and $10$ kOe, thus
implying that the applied $H$ field value cannot be at the origin of
the observed $T_{\textrm{N}}^{\textrm{Ce}}$ suppression. The
explanation possibly relies on the modification of the indirect RKKY
coupling through the involvement of FeAs bands into SC. Still, the
results are even more intriguing in view of the possible occurrence
of subtle many-body effects like the Kondo screening and its
interplay with SC. This issue has been discussed in the literature
from a theoretical point of view, showing how a competition between
these two mechanisms is at work in oxy-pnictides \cite{Pou08}. Our
results display a common enhancement of Ce magnetism and SC and are
not in contrast with such theoretical scenario.

Summarizing, in this Letter we discussed the effects of the in-plane
and out-of-plane chemical substitutions in CeFeAsO realized by
Fe$_{1-x}$Co$_{x}$ and O$_{1-x}$F$_{x}$ dilutions, respectively.
Remarkably, $\mu^{+}$SR enlightened that the spin-density-wave phase
is suppressed in a quantitatively identical fashion in both cases,
showing that the degree of the in-plane disorder is not playing any
substantial role. At variance with the behavior for the
spin-density-wave, such in-plane disorder strongly affects the
superconducting properties resulting in a much more effective
suppression of $T_{\textrm{c}}$ for the Fe$_{1-x}$Co$_{x}$ rather
than for the O$_{1-x}$F$_{x}$ dilution. The unusual behavior of Ce
magnetism was discussed in the light of the different kinds of
disorder introduced by the two considered chemical substitutions.
Our surprising results clearly deserve a more detailed computational
investigation concerning the precise role of indirect RKKY-like
coupling among the localized Ce$^{+}$ magnetic moments, also in
relation with the anomalous enhancement of the ordering temperature
$T_{\textrm{N}}^{\textrm{Ce}}$ in the presence of superconductivity.

G. Prando acknowledges support from the Leibniz-Deutscher
Akademischer Austauschdienst (DAAD) Post-Doc Fellowship Program and
stimulating discussions with A. Akbari, J. Geck and P. Thalmeier. S.
Sanna and P. Carretta acknowledge support from Fondazione Cariplo
(research grant No. $2011$-$0266$). G. Lamura acknowledges support
from the FP$7$ European project SUPER-IRON (grant agreement no.
$283204$). S. Wurmehl acknowledges support by the Deutsche
Forschungsgemeinschaft DFG under the Emmy-Noether programme (grant
no. WU$595$/$3$-$1$). This work has been supported by the Deutsche
Forschungsgemeinschaft DFG through the Priority Programme SPP$1458$
(grants no. BE$1749$/$13$ and no. GR$3330$/$2$) and the Graduate
School GRK $1621$.

\section{Supplemental Material}

\subsection{Synthesis of samples}

Polycrystalline samples of CeFe$_{1-x}$Co$_{x}$AsO ($0 \leq x \leq
0.2$) have been prepared by a two-step solid-state reaction
similarly to what is described in Ref. \cite{Kon09}. In the first
step, CeAs was prepared from Ce slugs (Chempur, $99.9$\%) and As
lumps (Alfa Aesar, $99.999$\%) reacting a stoichiometric ratio in an
evacuated quartz tube placed in a two-zone furnace. In the second
step, we used the resulting CeAs and mixed it with Fe (Alfa Aesar,
$99.998$\%), Fe$_{2}$O$_{3}$ (Chempur, $99.999$\%), and Co
(Goodfellow, $99.99$\%) in a stoichiometric ratio. All starting
materials were homogenized by grinding in a ball mill. The resulting
powders were pressed into pellets under Ar atmosphere using a
pressure of $20$ kN, and subsequently annealed in an evacuated
quartz tube in a two-step annealing process at $940^\circ$C for $8$
h and at $1150^\circ$C for $48$ h.

In order to confirm the single-phase character of the polycrystals,
powder x-ray diffraction was performed on a Huber Guinier camera (Co
K$_{\alpha}$ radiation). The samples were either pure or contained
small amounts of CeAs, Co$_{3}$O$_{4}$, Fe$_{3}$O$_{4}$ and/or FeAs.
The microstructure and the composition were examined by scanning
electron microscopy (XL$30$ Philipps, IN$400$) equipped with an
electron microprobe analyzer for semi-quantitative elemental
analysis using the wavelength dispersive x-ray mode. The analysis
showed a good agreement between the nominal and the measured Co
contents. Accordingly, the nominal content is used to label the
samples.

\subsection{Characterization of samples}

\subsubsection{Electrical transport}

\begin{figure}[t!]
\begin{center}
\includegraphics[scale=0.2]{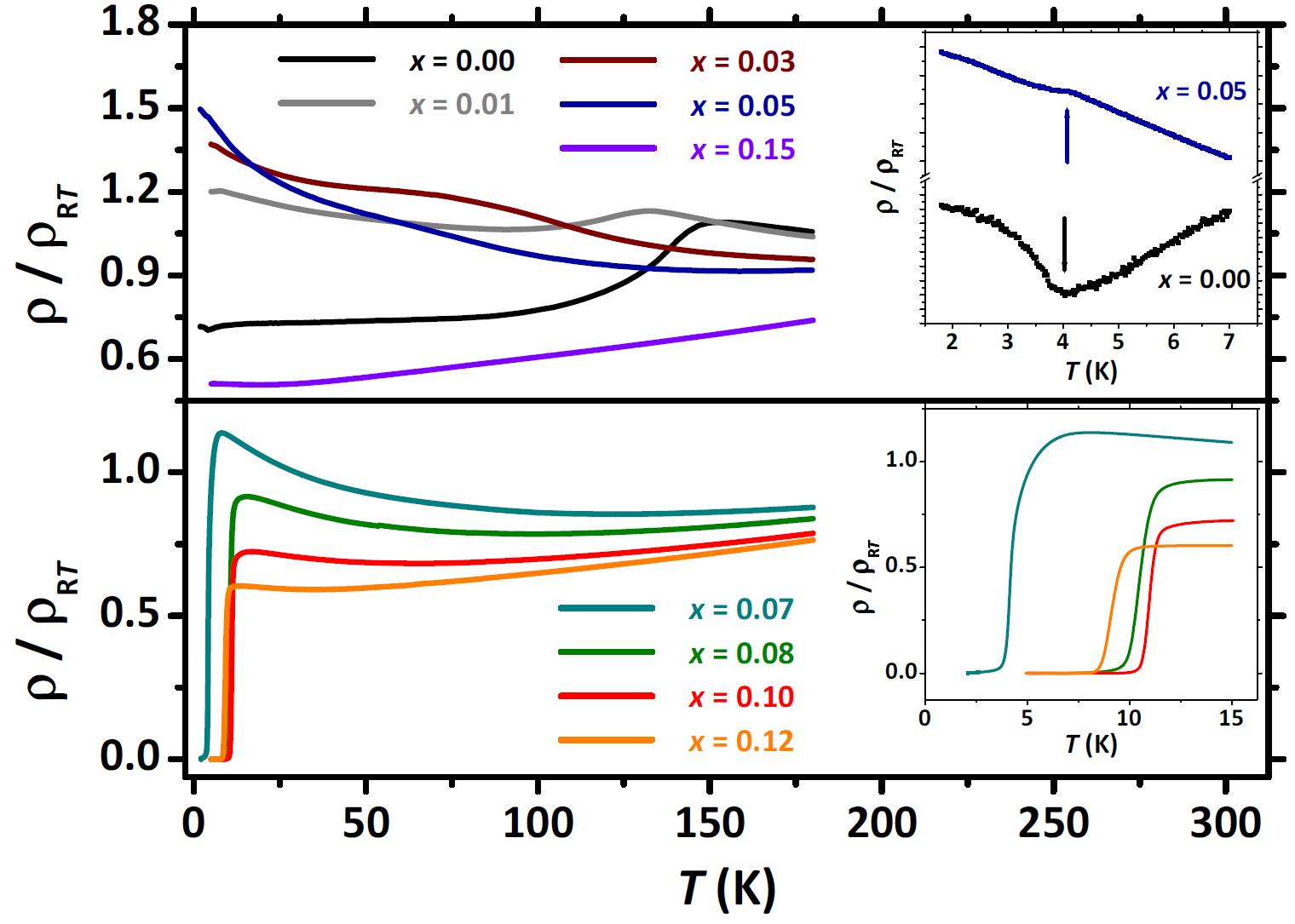}%
\caption{\label{GraResistivityVsX}(Color online) $\rho$ versus $T$
for CeFe$_{1-x}$Co$_{x}$AsO samples in ZF conditions. Data for
non-SC and SC samples are plotted in the upper and lower panels,
respectively. Upper inset: enlargement of data for non-SC samples
displaying kinks at $T_{\textrm{N}}^{\textrm{Ce}}$ (see arrow -
different scales are used for the two samples). Lower inset:
enlargement of data for SC samples in the low-$T$ region.}
\end{center}
\end{figure}
Measurements of electrical resistivity ($\rho$) were performed in
conditions of zero-magnetic field (ZF) as a function of temperature
($T$) by means of a standard four-probe setup. The $T$-dependence of
$\rho$ for CeFe$_{1-x}$Co$_{x}$AsO samples ($0 \leq x \leq 0.15$) is
shown in Fig.~\ref{GraResistivityVsX}. The curves are normalized
with respect to the relative $\rho_{\textrm{R}T}$ values at room $T$
(R$T$) for a better visualization of data. In the case of CeFeAsO,
the pronounced maximum at around $150$ K and the inflection point at
slightly lower $T$ can be related to the critical temperatures for
the structural and spin-density-wave (SDW) transitions,
$T_{\textrm{s}}$ and $T_{\textrm{N}}$ respectively
\cite{Hes09,Mae12}. These anomalies can be discerned up to $x =
0.03$ even if a sizeable and progressive broadening prevents one
from a precise definition of $T_{\textrm{s}}$ and $T_{\textrm{N}}$
for $x > 0.01$. However, a progressive suppression of both
$T_{\textrm{s}}$ and $T_{\textrm{N}}$ with increasing $x$ can be
inferred as a clear trend, in agreement with data reported in the
literature \cite{Sha13}. The clear kinks in $\rho$ versus $T$ curves
at $T \simeq 4$ K are associated to the magnetic ordering
temperature $T_{\textrm{N}}^{\textrm{Ce}}$ for the Ce sublattice
(see the upper inset of Fig.~\ref{GraResistivityVsX}). A further
increase in the level of Co-doping ($0.07 \leq x \leq 0.12$) leads
to the appearance of superconductivity (SC) with the highest value
for the critical temperature $T_{\textrm{c}}$ being realized for the
optimal-doping value $x \simeq 0.1$ (see the inset in the lower
panel of Fig.~\ref{GraResistivityVsX}) while SC is fully suppressed
by increasing the Co doping to values $x \geq 0.15$. Our data are in
good agreement with early reports in the literature
\cite{Sha13,Pra09,Zha10}.

\subsubsection{dc magnetometry}

\begin{figure}[t!]
\begin{center}
\includegraphics[scale=0.2]{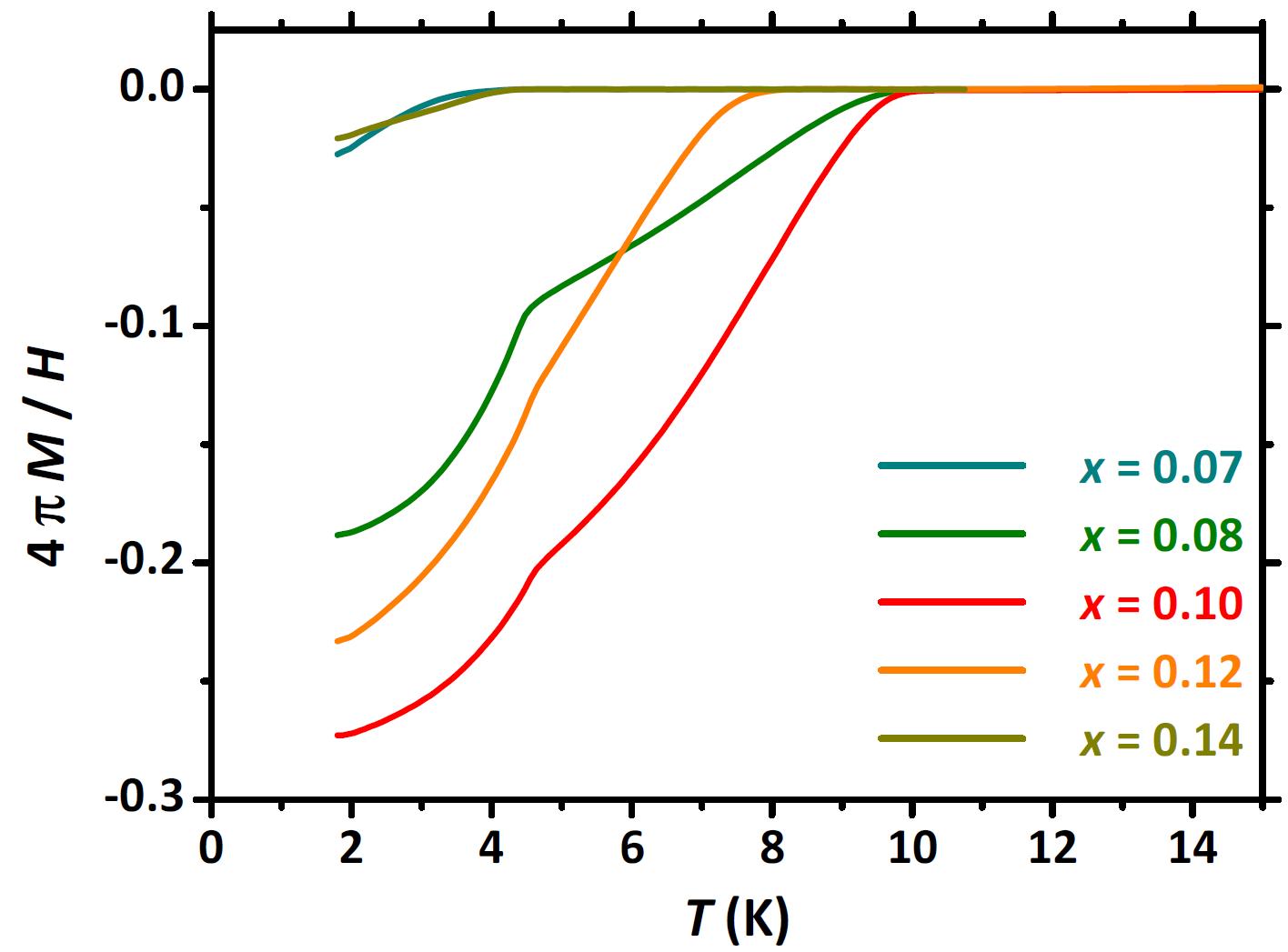}%
\caption{\label{CharacterizationSQUID}(Color online) dc
susceptibility for CeFe$_{1-x}$Co$_{x}$AsO SC samples. Measurements
were performed at $H = 20$ Oe in ZF-cooled conditions. The clear
kinks at $T \simeq 4$ K are associated with
$T_{\textrm{N}}^{\textrm{Ce}}$ (see also
Fig.~\ref{GraHDependenceCeOrdering}).}
\end{center}
\end{figure}
The magnetic characterization of the samples was performed by means
of a MPMS-XL$5$ SQUID dc magnetometer (Quantum Design). Results for
the SC materials are presented in Fig.~\ref{CharacterizationSQUID}
where a dome-like behaviour for the critical temperature
$T_{\textrm{c}}$ for the SC phase is evident. The maximum shielding
fraction can be roughly estimated as $30$\%, in good agreement with
previous reports on F-doped CeFeAsO \cite{Shi11,San10a} even if
systematically lower values are found in the current case (no
demagnetization effects from the geometrical shape of the grains
were considered at all). However, the picture is very useful in
order to check that the SC phase is not filamentary in the
considered materials even if a precise estimate of the SC fraction
can not be obtained from the current data.

\begin{figure}[t!]
\begin{center}
\includegraphics[scale=0.2]{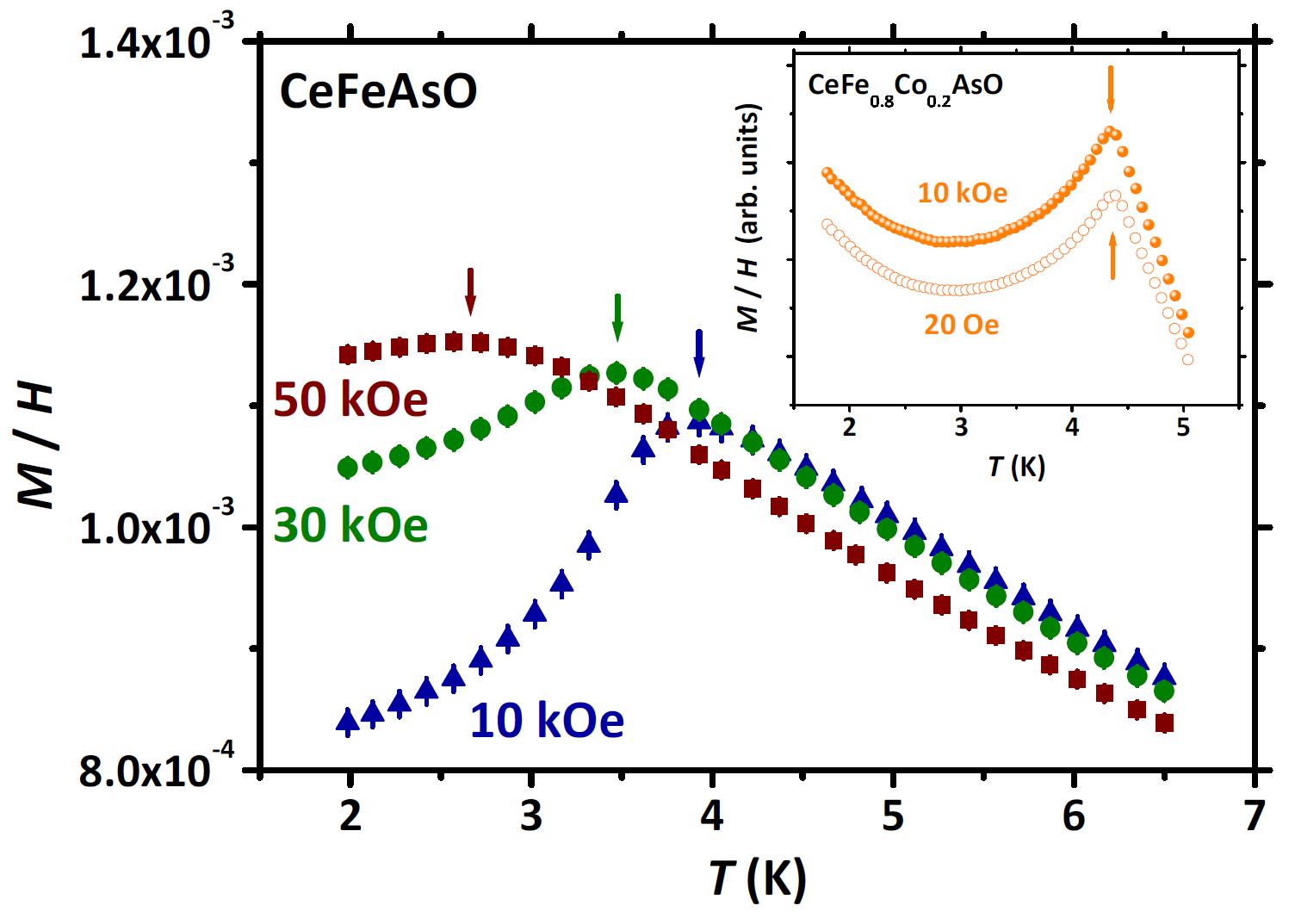}%
\caption{\label{GraHDependenceCeOrdering}(Color online) Main panel:
dc susceptibility for CeFeAsO in the high-$H$ regime. The arrows
indicate $T_{\textrm{N}}^{\textrm{Ce}}$. Inset: dc susceptibility
for CeFe$_{0.8}$Co$_{0.2}$AsO in the low-$H$ regime.}
\end{center}
\end{figure}

The value of $T_{\textrm{N}}^{\textrm{Ce}}$ can be clearly detected
by means of dc magnetometry since magnetization $M$ displays sharp
peaks in correspondence to those critical temperatures (see figs.
\ref{CharacterizationSQUID} and \ref{GraHDependenceCeOrdering}). A
strong dependence of such quantity on the external magnetic field is
detected as shown in the main panel of
Fig.~\ref{GraHDependenceCeOrdering}. However, it should be remarked
that in the limit $H \lesssim 10$ kOe the results display an almost
negligible dependence of $T_{\textrm{N}}^{\textrm{Ce}}$ on $H$ (see
the inset of Fig.~\ref{GraHDependenceCeOrdering}). The error bars
associated to results for SC samples at low-$H$ values are by far
too big if compared to what is obtained for non-SC materials (see
Fig.~\ref{CharacterizationSQUID}).

\subsection{Muon spin spectroscopy}

In a $\mu^{+}$SR experiment, a spin-polarized beam of positive muons
$\mu^{+}$ is implanted into the investigated sample. Coulomb
scattering processes typically lead to the thermalization of
$\mu^{+}$ at interstitial crystallographic sites without the loss of
spin polarization. After a mean life time $\tau_{\mu^{+}} \simeq
2.17 \mu$s, every $\mu^{+}$ decays emitting a positron $e^{+}$ with
linear momentum preferentially parallel to the direction of the spin
of the $\mu^{+}$ at the instant of the decay. As a consequence, a
spatially-resolved detection of $e^{+}$ as a function of time ($t$)
allows one to probe magnetism on a local scale \cite{Yao11}. The
typical output of a ZF-$\mu^{+}$SR experiment for magnetic materials
is the depolarization function
\begin{eqnarray}\label{EqGeneralFittingZF}
    \frac{A_{T}(t)}{A_{0}} & = &
    \left[1 - V_{\textrm{m}}(T)\right]
    e^{-\frac{\sigma_{\textrm{N}}^{2}
    t^{2}}{2}} {}\\ & & + \sum_{i = 1}^{N}
    \zeta_{i} \left[a_{i}^{\textrm{Tr}}(T) F_{i}(t)
    D_{i}^{\textrm{Tr}}(t) + a_{i}^{\textrm{L}}(T)
    D_{i}^{\textrm{L}}(t)\right] \nonumber
\end{eqnarray}
describing the $t$-dependence of the spin polarization of $\mu^{+}$
at $T$. Here, $A_{0}$ is an instrument-dependent parameter
corresponding to the condition of full spin polarization. In the
paramagnetic phase of the investigated material, $\mu^{+}$ are
subject to nuclear magnetism leading to a slow Gaussian damping
quantified by $\sigma_{\textrm{N}}$. In the presence of a magnetic
phase, a fraction $V_{\textrm{m}}(T)$ of $\mu^{+}$ probe a static
local magnetic field and this quantity reflects the magnetic volume
fraction of the sample at $T$, accordingly. The labels Tr and L in
eq. \eqref{EqGeneralFittingZF} refer to $\mu^{+}$ probing local
magnetic fields $B_{\mu}$ in transverse or longitudinal directions
with respect to the initial spin polarization, respectively. $F(t)$
oscillating functions represent the precession of $\mu^{+}$ around
the local field $B_{\mu}$ while $D^{\textrm{Tr},\textrm{L}}(t)$
quantify the damping of the signal. $D^{\textrm{Tr}}(t)$ is
typically associated with the static distribution of local magnetic
fields while $D^{\textrm{L}}(t)$ probes $1/\textrm{T}_{1}$-like
dynamical processes. The index $i$ in eq. \eqref{EqGeneralFittingZF}
allows for the possible presence of inequivalent crystallographic
sites for $\mu^{+}$ whose population is controlled by the parameters
$\zeta_{i}$ such that $\sum_{i} \zeta_{i} = 1$.

\begin{figure}[t!]
\begin{center}
\includegraphics[scale=0.2]{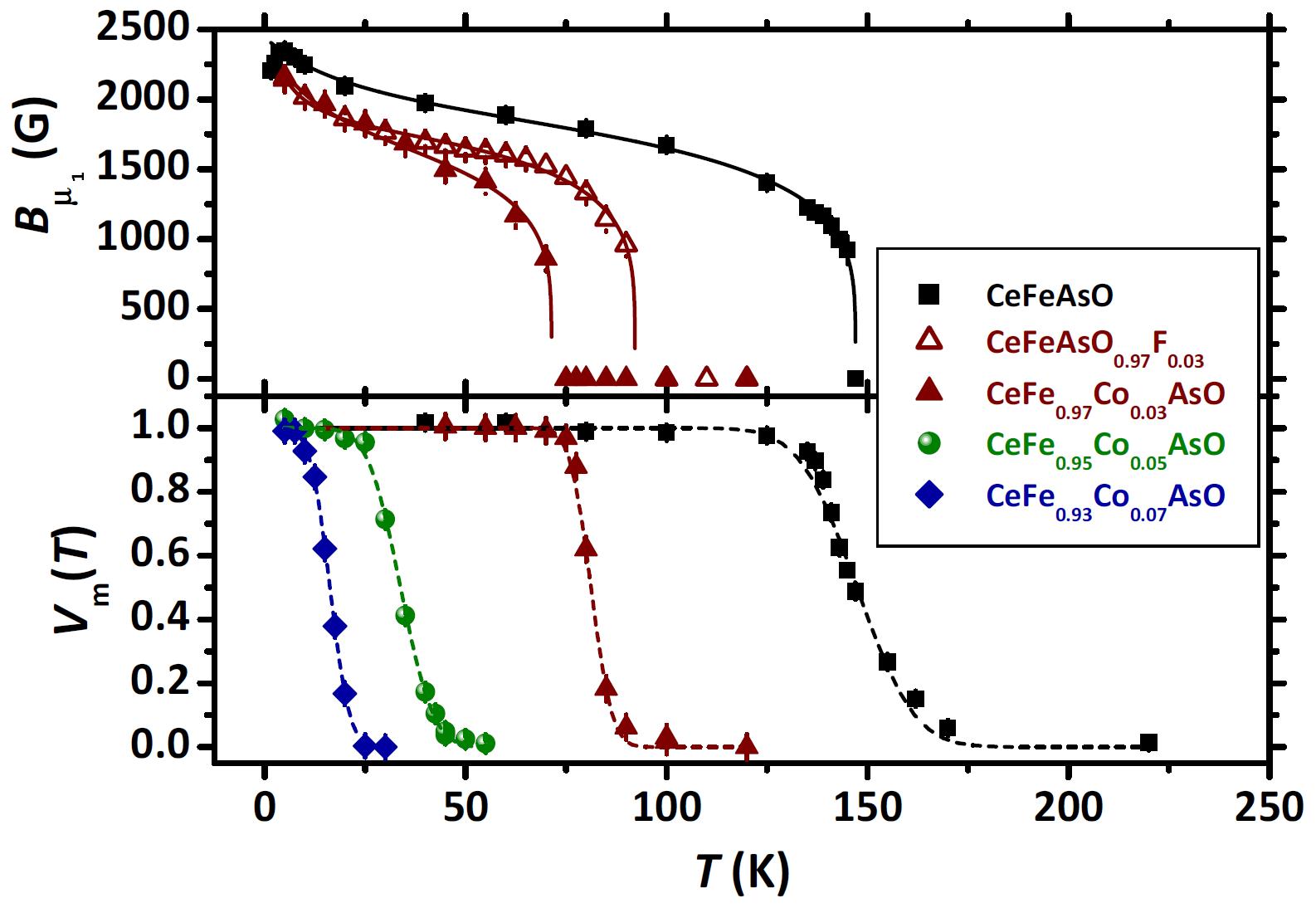}%
\caption{\label{MagnVolFractMuSR}(Color online) Upper panel:
$T$-dependence of the internal field at the $\mu^{+}$ site
$B_{\mu_{1}}$ for $x = 0$ and $x = 0.03$ (both F- and Co-dilutions
are considered). The continuous lines are best-fits to data
according to eq. \eqref{EqIntFieldMaeter}. Lower panel:
$T$-dependence of the magnetic volume fraction $V_{\textrm{m}}(T)$
for the CeFe$_{1-x}$Co$_{x}$AsO samples ($x \leq 0.07$). The dashed
lines are best-fits to data according to eq.
\eqref{EqMagneVolERFC}.}
\end{center}
\end{figure}
Measurements of $\mu^{+}$SR were performed at S$\mu$S (Paul Scherrer
Institute, Switzerland) on both the GPS and Dolly low-background
spectrometers at different temperatures ($T$) and in ZF conditions.
The presence of two inequivalent crystallographic sites for
$\mu^{+}$ is envisaged ($i = 1,2$) in agreement with previous
results on oxy-pnictides \cite{Mae09,DeR12,Pra13a}. In the current
case of CeFe$_{1-x}$Co$_{x}$AsO, one finds $\zeta_{1}/\zeta_{2}
\simeq 4$ for the occupation probabilities of the sites $\zeta_{i}$
independently from the actual $x$ value. The fitting results show
that the magnetic volume fraction $V_{\textrm{m}}(T)$ associated
with the SDW phase is maintained at around $100$\% in the low-$T$
saturation limit independently from $x$ (see
Fig.~\ref{MagnVolFractMuSR}, lower panel). $\mu^{+}$ implanted at
site $1$ probe long-range magnetism in CeFeAsO as clearly seen by
the long-lived coherent oscillation in the Tr component. In
particular, the function $F_{1}(t) = \cos\left(\gamma_{\mu}
B_{\mu_{1}} t\right)$ describes well the data while $F_{1}(t) =
J_{0}\left(\gamma_{\mu} B_{\mu_{1}} t\right)$ is needed in order to
reproduce the experimental trend for $x = 0.03$ ($J_{0}$ standing
for a zeroth-order first-kind Bessel function and $\gamma_{\mu}$
being the gyromagnetic ratio for $\mu^{+}$). This is a well-known
result for oxy-pnictide materials and it hints at the gradual
modification of the SDW phase from commensurability to
incommensurability with the underlying lattice upon increasing the
charge doping while at the same time reducing the nesting of the
Fermi surface \cite{Liu08,Car09}. The results for $B_{\mu_{1}}$ as a
function of $T$ are displayed in the upper panel of
Fig.~\ref{MagnVolFractMuSR}. In the same figure, it is shown how the
phenomenological function
\begin{equation}\label{EqIntFieldMaeter}
    B_{\mu_{1}}(T) = B_{\mu_{1}}^{\textrm{sat}}
    \left[1-\left(\frac{T}
    {T_{\textrm{N}}}\right)^{\alpha}\right]^{\beta}
    \left(1 + \frac{C_{\textrm{CW}}}{T-\theta}\right)
\end{equation}
first proposed in Ref. \cite{Mae09} correctly reproduce the
experimental data also in the case of the Co-doped $x = 0.03$
sample. In particular, the first term on the right side of eq.
\eqref{EqIntFieldMaeter} accounts for a double-exponent
power-law-like behaviour of the internal field due to the magnetic
ordering of Fe while the second term accounts for a feedback
paramagnetic polarization of the Ce sublattice induced by the
molecular field of Fe and governed by the Curie-Weiss constant
$C_{\textrm{CW}}$ and the Curie-Weiss temperature $\theta$
\cite{Mae09}. In the fitting procedure for all the three samples,
the values $\alpha = 2.1$ and $\beta = 0.2$ have been kept fixed in
order to reduce the number of free parameters (the values have been
extracted from what is reported in the literature \cite{Mae09}).

The gradual disordering of the magnetic phase leads to the
disappearance of oscillations in the Tr component for $x \geq 0.05$,
$F_{1}(t)$ being overdamped by a sizeable distribution of static
local magnetic fields. This corresponds to $F_{1}(t) = 1$ for both
$x = 0.05$ and $0.07$ samples with transversal dampings described as
$D_{1}^{\textrm{Tr}}(t) = \exp\left(-\lambda_{1}^{\textrm{Tr}}
t\right)$. At the lowest investigated temperature ($T = 5$ K), one
has $\lambda_{1}^{\textrm{Tr}} \simeq 45$ $\mu$s and $25$ $\mu$s for
$x = 0.05$ and $0.07$, respectively. Such values are quite typical
in the region of coexistence between magnetism and SC for
oxy-pnictides \cite{San10a,Shi11}. On the other hand, $\mu^{+}$
implanted at site $2$ probe sizeable distributions of local magnetic
fields for all the samples and $F_{2}(t) = 1$, accordingly, with
$D_{2}^{\textrm{Tr}}(t) = \exp\left(-\lambda_{2}^{\textrm{Tr}}
t\right)$. At $T = 5$ K, one has $\lambda_{2}^{\textrm{Tr}} \simeq
25$ $\mu$s, $20$ $\mu$s, $15$ $\mu$s and $8$ $\mu$s for $x = 0$,
$0.03$, $0.05$ and $0.07$, respectively.

$T_{\textrm{N}}$ is precisely quantified from the actual
$T$-dependence of the magnetic volume fraction $V_{\textrm{m}}(T)$
after a fitting procedure according to the error-function-like
expression \cite{San10a,Shi11,DeR12}
\begin{equation}\label{EqMagneVolERFC}
    V_{\textrm{m}}(T) = \frac{1}{2} \; \textrm{erfc}\left[\frac{T -
    T_{\textrm{N}}} {\sqrt{2}\Delta}\right]
\end{equation}
where the complementary error function $\textrm{erfc}(x)$ is defined
as
\begin{equation}
    \textrm{erfc}(x) = \frac{2}{\sqrt{\pi}} \int_{x}^{+\infty}
    e^{-t^{2}} dt.
\end{equation}
In particular, $T_{\textrm{N}}$ turns out to be defined as the $T$
value corresponding to $50$\% of the magnetic volume fraction.


\end{document}